# Quantifying the information transduction of biochemical reaction cascades: transcription of mRNA from ligand stimulation

T. Tsuruyama

Running title: Channel capacity of signal transduction

**ABSTRACT**

A cell has the ability to convert an environmental change into the expression of genetic information through a chain of intracellular signal transduction reactions. Here, we aimed to develop a method for quantifying this signal transduction. We showed that the channel capacities of individual steps in a given general model cascade were equivalent in an independent manner, and were given by the entropy production rate. Signal transduction was transmitted by fluctuation of the entropy production rate and quantified transduction was estimated by the work done in individual steps. If the individual step representing the modification to demodification of the signal molecules is considered to be a Szilard engine, the maximal work done is equivalent to the chemical potential change of the messenger that is consumed during the modification reaction. Our method was applicable to calculate the channel capacity of the MAPK cascade. In conclusion, our method is suitable for quantitative analyses of signal transduction.

**INTRODUCTION**

Several analytical methods applicable to dynamic biochemical networks and systems biology have been developed in the last few decades (1, 2). In particular, biological information transmission, i.e., signaling cascades, has received widespread interest. A signaling cascade is generally carried out by a chain reaction between proteins and messengers, such as adenosine triphosphate (ATP). In actuality, following an extracellular biochemical change, a set of proteins, including kinases, is serially phosphorylated and dephosphorylated in a cyclic manner with hydrolysis of ATP. A phosphorylated kinase has the ability to phosphorylate other kinases. Through this chain of phosphorylation or other modifications that are free energy change dependent, signal transduction may be described from the perspective of kinetics and thermodynamics in association with information theory.

There have been extensive studies regarding the relationship between thermodynamics and information processing. One such systematic discussion was initiated by Szilard (3). He argued that positive work can be extracted from an isothermal cycle, i.e., a Szilard engine, via Maxwell's demon, i.e., a feedback controller (4, 5). The fluctuation theorem (FT) is a key to understanding a biological signaling cascade by calculation of information transmitted by a Szilard engine. Some studies have verified the FT by analyses of cascades occurring in chemical reactions (6, 7) and molecular motors (8). A theoretical information thermodynamics model has been recently developed by Sagawa et al. in association with FT; the maximal work by the feedback controller was described by the information that was received by the controller (9-12). This conclusion was recently applied to evaluation of a chemotaxis model in Escherichia coli (13).

In the current study, we aimed to confirm the thermodynamic formulation of information based on a model of actual biochemical signaling cascades, such as cytokine-induced cascades (14-24), and to calculate the amount of signal transduction according to the entropy production rate. For this purpose, we applied FT to estimate the transduced signals. In addition, we examined whether the step cycle reaction between activation and inactivation of signaling molecules may be regarded as a Szilard engine to allow for calculation of the chemical work done. Finally, we aimed to quantify gene expression, i.e., transcription of messenger RNA (mRNA).

## MATERIALS AND METHODS

**Statistical analysis.** Regression analysis was performed using SPSS 10.0.5 (SPSS Inc.; Chicago, IL, USA).

## RESULTS

**Biological information transmission (signal transduction).** Let us consider a cell system as an open homogeneous reactor in contact with chemiostats of reactants and products, which drive the system out of equilibrium. The system is assumed to be isothermal and isovolumic. A model signaling cascade consisting of $n+1$ steps is then considered as follows (Fig. 1):

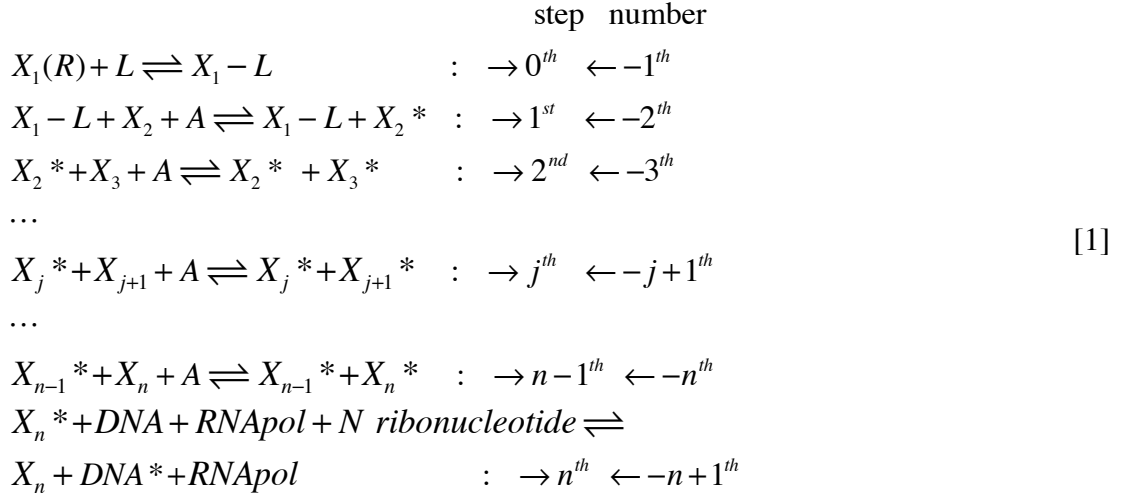

$$
\begin{aligned}
&X_1(R) + L \rightleftharpoons X_1 - L &&: \rightarrow 0^{th} \quad \leftarrow -1^{th} \\
&X_1 - L + X_2 + A \rightleftharpoons X_1 - L + X_2^* &&: \rightarrow 1^{st} \quad \leftarrow -2^{th} \\
&X_2^* + X_3 + A \rightleftharpoons X_2^* + X_3^* &&: \rightarrow 2^{nd} \quad \leftarrow -3^{th} \\
&\ldots \\
&X_j^* + X_{j+1} + A \rightleftharpoons X_j^* + X_{j+1}^* &&: \rightarrow j^{th} \quad \leftarrow -j+1^{th} \\
&\ldots \\
&X_{n-1}^* + X_n + A \rightleftharpoons X_{n-1}^* + X_n^* &&: \rightarrow n-1^{th} \quad \leftarrow -n^{th} \\
&X_n^* + DNA + RNApol + N \text{ ribonucleotide} \rightleftharpoons \\
&X_n + DNA^* + RNApol &&: \rightarrow n^{th} \quad \leftarrow -n+1^{th}
\end{aligned}
\quad [1]
$$

Each step is a modification/demodification of signaling molecules and is maintained by a chemical reservoir that provides a signaling messenger *A* such as ATP. $X_{\pm j}$ and $X_{\pm j}^*$ ($1 \leq j \leq n$) denote the signaling molecule non-modified and modified by the messenger, respectively. The right-pointing arrow in Eq. [1] represents one orientation of signal transduction and the left-pointing arrow indicates the opposite orientation. Here, the suffix plus "+" implies the duration of signaling from $X_j$ to $X$; and the suffix "-" implies the duration of signaling from $X_{j+1}$ to $X_j$.

The first reaction in the cascade is the uptake or binding of the extracellular molecule, the ligand (*L*), by $X_1$. This promotes the binding of *A* to $X_2$, the first intracellular signaling molecule, leading to modification of $X_2$ into $X_2^*$. The molecule $X_1$ is designated as the receptor (*R*) because it is located on the cell membrane to receive the external stimulation by *L*. $X_2^*$ has the ability to modify $X_3$ into $X_3^*$ by promoting the binding of *A* to $X_3$. The cascade continues in this manner, such that the $j^{th}$-signaling molecule $X_j$ induces the modification *of* $X_{j+1}$ into $X_{j+1}^*$. In the final $n^{th}$ step, the signaling molecule $X_n^*$ translocates to the cell nucleus and, after binding DNA in a sequence-specific manner, induces the transcription of mRNA from genes on DNA* by RNA polymerase (RNApol) to form a DNA-RNA complex. In individual steps, demodification of $X_j^*$ occurs spontaneously or via enzymatic reaction by a phosphatase or other demodification enzymes, and the pre-stimulation steady state is subsequently recovered. Thus, individual steps form cyclic reactions between modification and demodification. In summary, the above cascade in Eq. [1] describes a cascade of reactions from the reception of the extracellular stimulus to the induction of the transcription of mRNA. In this cascade, biological information is considered to be transmitted from the $0^{th}$ step to the $n^{th}$ step.

According to this cascade model, let us consider all of the possible distinct signaling cascades. The maximum channel capacity of a cascade of distinct signaling events during a given period will differ in the order of the symbols used and in the selection of $X_j$. The cell signaling formulation then begins by assuming a total continuous duration,

$$\tau \triangleq \sum_{j=1}^{n} X_{\pm j} \tau_{\pm j}. \qquad [2]$$

Considering the orientation of the signaling transmission, we assigned plus and minus values to $\tau_{+j}$ and $\tau_{-j}$, respectively. The total number of cascades or different combinations of such molecules can be computed. The total number of signaling molecules $X$ is expressed as

$$X \triangleq \sum_{j=1}^{n} X_{\pm j}, \qquad [3]$$

where $X_{\pm j}$ is the number of the $j^{th}$ signaling molecules in the cascade. A negative value is assigned to $\tau_{-j}$ (Figure 2). The probability of the $j^{th}$ molecule in the $j^{th}$ step and in the $-j^{th}$ step is given by:

$$p_j \triangleq \frac{X_{\pm j}}{X} \quad (j = 1, 2, \ldots n), \qquad [4]$$

$$X_j^* = X_{-j} \qquad [5]$$

Here, asterisk "*" denotes the active species that bind mediator $A$. In the $j^{th}$ step, the speciesd $X_j$ functions as the signaling molecule and the speciesd $X_{-j}$ functions as the signaling molecule in the $-j^{th}$ step. Further,

$$\sum_{j=1}^{n} p_{\pm j} = 1 \qquad [6]$$

$p_{\pm j}$ is the selection probability of the $\pm j^{th}$ step, and $\tau_{\pm j}$ signifies the duration of the step. Eq. [2] can be rewritten using Eqs. [3] and [6] so that

$$\tau = X \left( \sum_{j=1}^{n} p_j \tau_j + \sum_{j=1}^{n} p_{-j} \tau_{-j} \right). \qquad [7]$$

Here, we define the logarithm of the total Shannon entropy:

$$\log\psi = X\left(-\sum_{j=1}^{n} p_j \log p_j + \sum_{j=1}^{n} p_{-j} \log p_{-j}\right) \quad [8]$$

to identify the optimal coding system so that the Shannon entropy in a code is maximized under the condition that the cascade duration τ is constant. In above, we set the item $+p_{-j} \log p_{-j}$, because this item gives negative Shannon entropy in the given cascade in constrast to $-p_j \log p_j$ that gives the positive entropy. Here, we introduced arbitrary coefficients α and β with reference to Eqs. [6], [7], and [8].

$$d\log\psi - d\alpha \sum_{j=1}^{n} p_{\pm j} - \beta d\tau = 0 \quad [9]$$

From [7],

$$d\tau = X\left(\sum_{j=1}^{n} \tau_j dp_j + \sum_{j=1}^{n} \tau_{-j} dp_{-j}\right) + dX\left(\sum_{j=1}^{n} p_j \tau_j + p_{-j}\tau_j\right)$$

$$= X\left(\sum_{j=1}^{n} \tau_j dp_j - \sum_{j=1}^{n} \tau_{-j} dp_j\right) + dX\left(\sum_{j=1}^{n} p_j \tau_j + p_{-j}\tau_j\right) \quad [10]$$

To maximize the entropy under the condition that satisfies Eq. [9], taking the differential of [8] gives:

$$d\log\psi = -dX\sum_{j=1}^{n} p_j \log p_j + dX\sum_{j=1}^{n} p_{-j} \log p_{-j} - X\sum_{j=1}^{n}(2+\log p_j + \log p_{-j})dp_j \quad [11]$$

In above, we used $dp_{-j} = -dp_j$ from [4].

Substituting Eq. [10] and Eq. [11] into Eq. [9] gives:

$$-dX\left[\sum_{j=1}^{n} p_j \log p_j - \sum_{j=1}^{n} p_{-j} \log p_{-j} + \beta\sum_{j=1}^{n} p_j \tau_j + \beta\sum_{j=1}^{n} p_{-j}\tau_{-j}\right] +$$

$$\sum_{j=1}^{n} dp_j\left[-\alpha - \beta X\left(\tau_j - \tau_{-j}\right) + X(-2 - \log p_j - \log p_{-j})\right] = 0 \quad [12]$$

If $dX$ and $dp_j$ are treated as independent variables, then

$$\sum_{j=1}^{n} p_{\pm j} \log p_{\pm j} + \beta \sum_{j=1}^{n} p_{\pm j} \tau_{\pm j} = 0 \qquad [13]$$

$$-\alpha - \beta X\left(\tau_j - \tau_{-j}\right) + X\left(-2 - \log p_j - \log p_{-j}\right) = 0 . \qquad [14]$$

To satisfy Eq. [13] and [14], we have:

$$\alpha = -2X . \qquad [15]$$

and

$$-\log p_j = \beta \tau_j . \qquad [16]$$

$$-\log p_{-j} = -\beta \tau_{-j} \qquad [17]$$

**The average entropy production rate at the individual steps in the cascade.**
Subsequently, we defined $p(j+1|\ j)$, the conditional probability of the $j + 1^{th}$ step given the $j^{th}$ step, and $v(j+1|\ j)$, the transitional rate for the $j^{th}$ step to the $j + 1^{th}$ step in the same orientation of signaling. On the other hand, we defined $p(-j\ |\ -j+1)$, the conditional probability of the $-j^{th}$ step given the $-j+1^{th}$ step, and $p(-j\ |\ -j+1)$, the transitional rate for the $-j + 1^{th}$ step to the $-j^{th}$ step in the opposite orientation of signaling in the given cascade. The conditional probabilities $p(j+1\ |\ j)$ and $p(-j\ |\ -j+1)$ are equivalent to the transitional probability that the signaling events proceed from the $j^{th}$ step to the $j + 1^{th}$ step and from the $-j + 1^{th}$ step to the $-j^{th}$ step, respectively, when the probability is determined by the immediate preceding step. We considered that the cell system stays balanced around the steady state, as follows:

$$p(j+1|j)v(j+1|j) = p(-j|-j+1)v(-j|-j+1) .$$

$$[18]$$

Therefore, we have:

$$\log \frac{p(-j|-j+1)}{p(j+1|j)} = \log \frac{v(j+1|j)}{v(-j|-j+1)} . \qquad [19]$$

Bayes' theorem gives:

$$\frac{p(j+1|j)}{p(-j|-j+1)} = \frac{p_j}{p_{-j+1}} \qquad [20]$$

Therefore, we have:

$$\log \frac{p(j+1|j)}{p(-j|-j+1)} = \log \frac{p_j}{p_{-j+1}} \tag{21}$$

Using Eq. [16],

$$\frac{1}{\tau_j - \tau_{-j+1}} \log \frac{p(j+1|j)}{p(-j|-j+1)} = -\beta \frac{\tau_j + \tau_{-j+1}}{\tau_j - \tau_{-j+1}} \tag{22}$$

In common time course of the $j^{th}$ to $j + 1^{th}$ step in many reports (14–24), $-\tau_{-j} \gg \tau_j$ and the right side of [22] gives:

$$\frac{1}{\tau_j - \tau_{-j+1}} \log \frac{p(j+1|j)}{p(-j|-j+1)} \simeq \beta \tag{23}$$

Here, we defined the average entropy production rate during transduction duration $t_j = \tau_j - \tau_{-j+1}$ using an arbitrary parameter $\Delta \zeta_j(s)$ that represents entropy production rate change during the $j^{th}$ step:

$$\Delta \bar{\zeta}_j \triangleq \frac{1}{\tau_j - \tau_{-j+1}} \int_0^{\tau_j - \tau_{-j+1}} \Delta \zeta_j(s) ds \tag{24}$$

Then we have from FT and Eq. [22]:

$$\frac{1}{\tau_j - \tau_{-j+1}} \log \frac{p(j+1|j)}{p(-j|-j+1)} = \Delta \bar{\zeta}_j \tag{25}$$

From Eq. [22] and Eq. [25],

$$\beta = \Delta \bar{\zeta}_j \triangleq \Delta \zeta \tag{26}$$

Therefore, we obtained an important result that the average entropy production rate is equivalent at the individual steps in the cascade. Using Eq. [16],

$$-\log p_j = \tau_j \Delta \zeta \quad (\tau_j > 0) \tag{27}$$

$$\log p_{-j} = \tau_{-j} \Delta \zeta \quad (\tau_{-j} < 0) \tag{28}$$

Here, $\tau_{-j}$ has a negative value.

$$\log \psi = X \left( -\sum_{j=1}^{n} p_j \log p_j + \sum_{j=1}^{n} p_{-j} \log p_{-j} \right) = X \Delta \zeta \left( \sum_{j=1}^{n} p_j \tau_j + \sum_{j=1}^{n} p_{-j} \tau_j \right) = \tau \Delta \zeta \tag{29}$$

Channel capacity of the whole cascade is given by

$$C = \lim_{x \to \infty} \frac{\log \psi}{\tau} = \Delta \zeta \qquad [30]$$

**Mutual transformation in the model cascade.** Here, we considered the mutual information in actual cascade. In an actual cascade as depicted in Fig. 3, activation of the *j+1*-th signal is delayed following activation of the *j*-th step: when modification of $X_j$ occurs, modification of $X_{j+1}$ is not observed immediately. To an observer, modification of $X_j$ comes with the error probability $e_j$ that $X_{j+1}$ is unmodified, and the probability $(1-e_j)$ that $X_{j+1}$ is modified. Here we set $\xi_j = [e_j \log e_j + (1-e_j) \log(1-e_j)]$; $0 \geq \xi_j \geq -1$. The entropy $H_j$ and the conditional entropies $H_j(X_j | X_{j+1})$ are given by:

$$H_j = -p_j \log p_j + p_{-j} \log p_{-j} \qquad [31]$$

$$H(j|j+1) = \xi_j p_{-j} \qquad [32]$$

To choose $p_j$ and $p_{-j}$ in such a way as to maximize them, we introduce the following function $U_j$ using the undetermined parameter $\lambda$:

$$U_j = -p_j \log p_j + p_{-j} \log p_{-j} + \xi_j p_{-j} + \lambda (p_j - p_{-j}) \qquad [33]$$

Then

$$\frac{\partial}{\partial p_j} U_j = -1 - \log p_j + \lambda \qquad [34]$$

$$\frac{\partial}{\partial p_{-j}} U_j = 1 + \log p_{-j} - \xi_j - \lambda \qquad [35]$$

Setting the right-hand parts of [29] and [30] to zero and eliminating $\lambda$:

$$-\log p_j + \log p_{-j} = \xi_j \qquad [36]$$

$$\therefore \log \frac{p_j}{p_{-j}} = -\xi_j = t_j \Delta \zeta \qquad [37]$$

From [27] and [28], we have:

$$p_j = \frac{p_j^0}{1 + \exp \xi_j}, \quad p_{-j} = \frac{\exp \xi_j}{1 + \exp \xi_j} p_j^0 \qquad [38]$$

Channel capacity of the jth step $C_j$ is given by:

$$C = \max\left[H_j(X_j) - H_j(X_j | X_{j+1})\right] = \max\left[\frac{\exp t_j \Delta\zeta - 1}{\exp t_j \Delta\zeta + 1} t_j \Delta\zeta\right] \simeq t_j \Delta\zeta \quad (t_j \to \infty) \quad [39]$$

In above we set the duration sufficiently long according to previous reports (14–24).

**Biochemical signaling as a Szilard engine.** Next, let us consider the possibility that the biochemical signaling cascade is an example of a Szilard engine. Suppose that the cell signaling system is divided into $n$ number of rooms, which correspond to an individual $j^{th}$ room ($1 \leq j \leq n$) as follows. Each room has all of both $X_j^*$ and $X_j$ species ($1 \leq j \leq n$), whose concentrations are identical to $X_j^{*st}$ and $X_j^{st}$ at the steady state when the stimulus is absent.

(i) The feedback controller, i.e., Maxwell's demon (25), observes whether the step in the $j^{th}$ room ($j^{th}$ step) proceeds in the same or opposite direction of signaling and measures the change in the concentration of the messenger $A_j$ in the $j^{th}$ room.

(ii) If the $j^{th}$ ($1 \leq j \leq n$) step proceeds in the same direction of signaling, $A_j$ decreases from $A_{ij}$ to $A_{fj}$. Then, the controller introduces $\Delta X_{J+1}^*$ to the $j + 1^{th}$ room from the $j^{th}$ room and induces $\Delta X_{J+1}$ to the $j^{th}$ room from the $j + 1^{th}$ room. In this case, $I_j > 0$. In contrast, the $j^{th}$ step proceeds in the opposite direction of signaling, $A_j$ increases from $A_{ij}$ to $A_{fj}$. Then the controller introduces $\Delta X_{j+1}$ to the $j + 1^{th}$ room from the $j^{th}$ room and induces $\Delta X_{J+1}^*$ from the $j + 1^{th}$ room to the $j^{th}$ room. In this case, $I_j < 0$.

(iii) After the exchange between $\Delta X_{J+1}$ and $\Delta X_{J+1}^*$ in (ii) by the feedback controller is complete, the controller lets the system recover to the initial state.

The above conditions (i)–(iii) form a cycle consisting of chemical reactions (Fig. 4). When the controls in (ii) are quasi-static and the individual step recovers to the initial steady state, the cyclic reaction satisfies the Szilard engine. Thus, the given reaction cascade in Eq. [1] can be regarded as a Szilard engine that may play fundamental roles similar to a Carnot cycle in a classical heat engine.

**Chemical work in the model cascade.** Here, let us consider the biological signaling cascade as a kind of chemical Szilard engine. Setting the difference in the Helmholtz free energy from the initial state $F_{ij}$ to $F_{fj}$, as $\Delta F_j \equiv F_{fj} - F_{ij}$,

$$w_{a,j} \leq -\Delta F_j + k_B T H_j(X_j; X_{j+1}) \quad [40]$$

Here, $k_B$ is the Boltzmann constant and $T$ is the temperature of the cell system. In actuality, a biological system is kept in an isovolumic and isothermal state and we designated advection work of signaling molecules, $w_{a,j}$ by $j+1^{th}$ room or $j-1^{th}$ room to $j^{th}$ room; therefore, $\Delta F = 0$ in Eq. [40]. Thus, we can obtain using Eq.[40]:

$$w_{a,j} \leq k_B T H_j(X_j; X_{j+1}) \qquad [41]$$

The equality of works $w_{a,j}$ ($0 \leq j \leq n-1$) was obtained from Eq. [39] and [41], because the feedback system in Figure 4 is an Szilard engine:

$$w_{a,j} = k_B T H_j(X_j; X_{j+1}) = k_B T \tau_j \Delta \zeta \qquad [42]$$

**Kinetics of signal transduction.** Next, let us consider an actual cascade model. In a biological system, signaling occurs continuously at a steady homeostatic state. Stimulation of this stable system produces a transient fluctuation $\Delta X_j *(0 \leq j \leq n-1)$ that actually carries the signal and results in an increase in a specified mRNA. Biological information is transmitted through a change in fluctuation. Let us consider quantifying the information transduction in a biological system that employs fluctuation (14–24). The kinetic rate of the reaction along with the signal transduction is given by:

$$v(j|-j+1) = k_j A X_j * X_{j+1} \qquad [43]$$

By contrast, the kinetic rate of the demodification reaction opposite to the signal transduction is given by:

$$v(j+1|j) = k_{-j} X_j * X_{j+1} * \qquad [44]$$

Here, $k_j$ and $k_{-j}$ represent the kinetic coefficients. Using Eqs. [19], [43] and [44], we obtain:

$$\log \frac{p(-j|-j+1)}{p(j+1|j)} = \log \frac{k_j A X_{j+1}}{k_{-j} X_{j+1} *} \qquad [45]$$

Here, we note the concentration of the signaling molecules at the homeostatic state using the superscript $st$ and the concentration fluctuation using the symbol $\Delta$.

$$X_{j+1} = X_{j+1}^{st} + \Delta X_{j+1} \qquad [46]$$

$$X_{j+1}* = X_{j+1}^{st}* + \Delta X_{j+1}* \qquad [47]$$

Since the sum of the inactive form $X_j$ and the active form $X_j*$ is constant,

$$X_{j+1} + X_{j+1}* = X_{j+1}^0. \qquad [48]$$

then,

$$\Delta X_{j+1} + \Delta X_{j+1}^* = 0.  \quad [49]$$

Further, at the steady state, Eqs. [18], [43], [44] and [48] give the concentrations:

$$X_{j+1}^{st} = \frac{k_{-j} p(-j|-j+1)}{k_{-j} p(-j|-j+1) + k_j p(j+1|j) A} X_{j+1}^0. \quad [50]$$

$$X_{j+1}^{st*} = \frac{k_j p(j+1|j) A}{k_{-j} p(-j|-j+1) + k_j A p(j+1|j)} X_{j+1}^0. \quad [51]$$

The symbol $A$ stands for the initial concentration of the messenger.

$$\frac{1}{\tau_j - \tau_{-j+1}} \log \frac{k_j A X_{j+1}^{st}}{k_{-j} X_{j+1}^{st*}} \triangleq \overline{\zeta}_j. \quad [52]$$

For simplicity, we define $t_j = \tau_j - \tau_{-j+1}$. $t_j$ gives the total duration for $j$-step as shown in Figure 2. Using Eq. [52] and by introduction of the fluctuation of signaling molecules,

$$\log \frac{k_j A (X_{j+1}^{st} + \Delta X_{j+1})}{k_{-j} (X_{j+1}^{st*} + \Delta X_{j+1}^*)} = \overline{\zeta}_j t_j + \log \frac{1 + \Delta X_{j+1}/X_{j+1}}{1 + \Delta X_{j+1}^*/X_{j+1}^{st*}}. \quad [53]$$

The first item of the right-hand term is the entropy production in the $j^{th}$ step to the $j+1^{th}$ step at the steady state or pre-stimulation state, so that fluctuation of the entropy production rate occurs in the $j^{th}$ step to the $j+1^{th}$ step. From the FT, Eqs. [25], [27] and [45], we can obtain the fluctuation of entropy production rate:

$$\Delta \zeta = \frac{1}{t_j} \log \frac{1 + \Delta X_{j+1}/X_{j+1}}{1 + \Delta X_{j+1}^*/X_{j+1}^{st*}}. \quad [54]$$

Further, using the approximation $\log(1+x) \sim x$ in the logarithm item on the left side gives the following:

$$\Delta \zeta = -\frac{1}{t_j} \left( \Delta X_{j+1}/X_{j+1} - \Delta X_{j+1}^*/X_{j+1}^{st*} \right) = -\frac{1}{t_j} \Delta X_{j+1}^* \frac{X_{j+1}^0}{X_{j+1}^{st*} X_{j+1}^{st}}. \quad [55]$$

Using an integral form, we can describe the fluctuation of entropy production rate as:

$$\Delta \zeta = \frac{1}{\tau_j - \tau_{-j+1}} \int_0^{\tau_j - \tau_{-j+1}} \frac{X_{j+1}^0}{X_{j+1}^{st*} X_{j+1}^{st}} \frac{\Delta X_{j+1}^*}{\Delta t} dt = -\frac{1}{t_j} \int_0^{t_j} \frac{X_{j+1}^0}{X_{j+1}^{st*} X_{j+1}^{st}} \frac{\Delta X_{j+1}^*}{\Delta t} dt. \quad [56]$$

Then, we have

$$\Delta \zeta = -\frac{1}{t_j}\left[\log A\right]_{A_{ij}}^{A_{fj}} = -\frac{1}{t_j}\log\frac{A_{fj}}{A_{ij}} \quad .$$  [57]

Here, we used

$$\frac{dX_{j+1}^*}{dA} = \frac{1}{A} X_{j+1}^{st} * X_{j+1}^{st} \frac{1}{X_{j+1}^0} \quad .$$  [58]

$A_{fj}$ and $A_{ij}$ signify the concentration of the messenger $A$ at the initial and final state at the $j^{th}$ step, respectively. The external work by $j+1^{th}$ room or $j-1^{th}$ room to $j^{th}$ room can be estimated:

$$w_{ex}^j \leq k_B T t_j \overline{\Delta \zeta} = -k_B T \log\frac{A_{fj}}{A_{ij}}$$  [59]

Chemical potential of messenger $A$ is defined as:

$$\mu_{ATP}^j = \mu_{ATP}^0 + k_B T \log A_{ij}$$  [60]

and we can rewrite [59]:

$$w_{ex}^j \leq -\frac{\partial \mu_{ATP}^j}{\partial A}\Delta A\bigg|_{A_{ij}}$$  [61]

Using formula [59], let us consider the initial step of ligand binding ($0^{th}$ step). For the initial step in which the ligand $L$ binds to the receptor $X_1$,

$$w_{ex}^0 \leq -\frac{\partial \mu_L^0}{\partial L}\left(L_f - L_i\right) \quad .$$  [62]

The suffix "0" represents the initial step of the signaling cascade outside the cell. $L_f$ and $L_i$ signify the concentration of the ligand $L$ at the initial and final state at the $0^{th}$ step. In addition, let us consider the final step of gene expression, i.e., mRNA production ($n^{th}$ step). Biological signaling is completed by this transcription of mRNA in the nucleus. Transcription is the final signal transduction step in the production of an "informational" substance. Increased production of a specific mRNA is determined by the species $X_n$ binding to a unique DNA sequence in the nucleus of a cell. Such a sequence is named the binding motif. For instance, cytokine IL7 stimulation of a cell induces sequential phosphorylation of Janus kinase and STAT5, which binds to TCCNNGGA and similar palindromic motifs in target genes (26). Further, for the final $n^{th}$ step in which the $X_n^*$ binds to DNA(= $X_{n+1}$) in Eq. [1], using the concentration of ribonucleotide that is synthesized into mRNA, $\phi$:

$$\Delta \bar{\zeta}_n = \frac{1}{t_n} \int_0^{t_n} \frac{DNA^0}{DNA^{st} * DNA^{st}} \frac{\Delta DNA^{st*}}{\Delta t} dt$$

$$= \frac{1}{t_n} \int_0^{t_n} \frac{DNA^0_{j+1}}{DNA^{st} * DNA^{st}} \frac{\Delta DNA^{st*}}{\Delta t} \frac{d\phi}{dt} dt = \frac{1}{t_n} \log \frac{\phi_f}{\phi_i} ,$$

[63]

where $\phi_f$ and $\phi_i$ signify the ribonucleotide concentrations participating in the transcriptional process at the initial and final states of the $n^{th}$ step, respectively. $DNA$ signifies the concentration of total DNA. $DNA^{*st}$ and $DNA^{st}$ signify the concentration of the DNA-RNA complex and DNA at the steady state, respectively. In addition, these satisfy:

$$DNA^{st} + DNA^{*st} = DNA^0 .$$

[64]

From Eq. [63], we have

$$w^n_{a,j} \leq k_B T t_n \Delta \bar{\zeta}_n = -k_B T \log \frac{\phi_f}{\phi_i}$$

When the $r$ number of mRNA molecules consisting of $N$ ribonucleotides is produced in the transcriptional process,

$$\phi_f = \phi_i - rN$$

[65]

Then we have:

$$w^n_{a,j} \leq -k_B T \log \frac{\phi_i - rN}{\phi_i} \simeq k_B T \frac{rN}{\phi_i}$$

[66]

In the above induction, we approximated that the concentration change of ribonucleotides in the nucleus, $rN$, is sufficiently small. mRNA actually consists of four types of ribonucleotides: adenine, uracil, guanine, and cytosine. The carried information, i.e., Shannon entropy $I_{RNA}$, by the $r$ number of mRNAs in Eq. [66] is given by:

$$I_{RNA} = -rN \left( p_G \log p_G + p_A \log p_A + p_U \log p_U + p_C \log p_C \right) = \theta rN$$

[67]

where $p_G$, $p_A$, $p_U$, and $p_C$ represent the frequencies in the mRNA of guanosine, adenosine, uridine, and cytidine, respectively. $\theta$ represents the average Shannon entropy per one ribonucleotide. For instance, when the nucleotide frequency is independent of neighboring nucleotides, $\theta$ is equivalent to 2. Using Eqs.[66], and [67], we have:

$$w^n_{ea} \leq k_B T \frac{I_{RNA}}{\phi_i \theta} = k_B T \frac{I_{RNA}}{I_{nuc}}$$

[68]

and

$$I_{nuc} \triangleq \phi_i \theta \quad [69]$$

Here, $I_{nuc}$ is total Shannon entropy, which is assigned to the nucleotide component of mRNA. In summary, for the cytoplasmic cascade,

$$w_{a,j} = -k_B T \log \frac{A_{fj}}{A_{ij}} \quad [70]$$

For the ligand-receptor interaction at the membrane,

$$w_{a,0} = -k_B T \log \frac{L_{fj}}{L_{ij}} \quad [71]$$

For the intranuclear interaction,

$$w_{a,n} = -k_B T \log \frac{\phi_f}{\phi_i} \quad [72]$$

The sum of the work of $j+1^{th}$ room for $j^{th}$ room, i.e., the whole chemical work can be obtained by

$$w = \sum_{j=1}^{n-1} p_j w_{a,j} = k_B T \sum_{j=1}^{n-1} p_{\pm j} \Delta \zeta \tau_{\pm j}$$
$$= k_B T \tau \Delta \bar{\zeta} = -k_B T \log \frac{A_{fn-1}}{A_{i1}} = -\frac{\partial \mu_{ATP}}{\partial A} \Delta A \quad [73]$$

Here, the chemical potential for the whole cascade is defined by:

$$\mu_{ATP}^t = \mu_{ATP}^0 + k_B T \log A \quad [74]$$

The above calculation used

$$\bar{\zeta} = -\frac{1}{\tau} \log \frac{A_{fn-1}}{A_{i1}} = \frac{1}{k_B T \tau} \frac{\partial \mu_{ATP}}{\partial A} \Delta A \quad [75]$$

**The channel capacity and information density of the MAPK cascade.** Signaling cascades have been studied extensively using models of MAPK pathways, in which the epidermal growth factor receptor, c-Raf, MAP kinase–extracellular signal-regulated kinase (MEK), and kinase–extracellular signal-regulated kinase (ERK) are phosphorylated constitutively following treatment with cytokines or other reagents. Ras-c-Raf- ERK cascade (RRE) is a ubiquitous signaling pathway that conveys mitogenic and differentiation signals from the cell membrane to the nucleus(14–24). We aimed to validate the result from Eq. [27] by adaptation of Eq. [54] into an actual signaling cascade using previously reported data. As an example, we analyzed the data from Petropavlovskaia *et al*. regarding activation of cascades in RIN-m5F rat islet cells stimulated by two types of ligands, full-length recombinant islet neogenesis-associated protein (rINGAP) and a 15 amino-acid fragment of INGAP-P(19). The RRE cascade was analyzed. In detail,

$$P + R \leftrightarrow R^*, R^* + Ras \leftrightarrow R^* + Ras^* \quad (S_1),$$
$$Ras^* + c-Raf \leftrightarrow c-Raf^*(S_2) + Ras^*,$$
$$c-Raf^* + MEK \leftrightarrow c-Raf^* + MEK^*,$$
$$MEK^* + ERK \leftrightarrow MEK + ERK^*(S_3)$$

[76]

Here, * represents activated signaling molecules. $P$ and $R$ represent the peptide ligands and their receptor, respectively. $M$ represents a signal mediator. We plotted below for durations $t_j$ ($j$ = 1, 2, and 3 for Ras, c-Raf, ERK, respectively):

$$\log \frac{1 + \Delta X_{j+1} / X_{j+1}}{1 + \Delta X_{j+1}^* / X_{j+1}^{st}{}^*} \sim -\log\left(1 + \Delta X_{j+1}^* / X_{j+1}^{st}{}^*\right)$$

[77]

In the above approximation, $\Delta X_{j+1}$ is small according to the experimental data (31-42). It was possible to perform regression analyses to calculate the channel capacity $\Delta \zeta$ in Eq. [54] from the gradient of the regression line as follows:

$$t_j = -\frac{1}{\Delta \zeta} \log\left(1 + \Delta X_{j+1}^* / X_{j+1}^{st}{}^*\right)$$

[78]

As a result, the RRE induced by INGAP-P was verified to function in a minimally redundant fashion ($n$ = 9, $P$ = 1.7 × 10$^{-3}$, $\Delta \zeta$ = 2.4× 10$^{-3}$ bit/min, Fig. 5A). In contrast, for rINGAP, regression analysis showed no significant correlation between $-\log\left(1 + \Delta X_{j+1}^* / X_{j+1}^{st}{}^*\right)$ and duration (Fig. 4B, Supplementary table 1).

## Discussion

In the current study, we discussed how a cell system transduces a signal on the basis of the minimum hypothesis. The hypothesis states that (i) the channel capacity is maximized for most signals within a certain duration; (ii) the signal is carried by the fluctuation around the steady state. This hypothesis includes a theoretical basis that can introduce a duration parameter to analyze the development of signal transduction over time.

Quantification of signal transduction from thermodynamic information was recently achieved by Sagawa and Ito (10-12). The aim of the current study was to apply this approach to the cascade reaction model in a complicated *in vivo* reaction. We discussed the signal transmission that is mediated by a number of complicated reaction networks. Our previous study reported the probability of phase transition and oscillation when the concentration of messenger was consistent with the non-linear kinetic model.

Eqs. [70], [71], and [72] give chemical work for exchange of signaling molecules $X_j$ and $X_j^*$, which is equivalent to the average entropy production of ligand-receptor interaction, cytoplasmic cascade reaction, and mRNA transcription in a cell system considered as a Szilard engine. In particular, we obtained a formula of the chemical potential change for an intracytoplasmic reaction cascade that is given only by the chemical potential of the messenger. This conclusion verified from the information thermodynamics that the messenger is defined as the signaling molecule. Then we obtained the channel capacity in the ligand-receptor interaction during the $\tau_0$:

$$C_0 = \bar{\zeta}_0 \tau_0 = \frac{1}{k_B T} \frac{\partial \mu_L}{\partial L} \Delta L \quad [79]$$

In the above, $\mu_L$ denotes the chemical potential of the ligand. The channel capacity in the cytoplasmic cascade during the $\tau$ from Eq. [75]:

$$C_j = \bar{\zeta} \tau = \frac{1}{k_B T} \frac{\partial \mu_{ATP}}{\partial A} \Delta A \quad (1 \leq j \leq n-1) \quad [80]$$

Likewise, the channel capacity in the mRNA synthesis is given by

$$C_n = \bar{\zeta}_n \tau_n = \frac{1}{k_B T} \frac{\partial \mu_{nuc}}{\partial nuc} \Delta nuc \quad [81]$$

In the above, $\mu_{nuc}$ denotes the chemical potential of the nucleotide. *nuc* denotes the concentration of the nucleotide. In conclusion, our formulation provides a theoretical basis of signal transduction. Although verification of our model analysis is technically difficult, we will report experimental application of the model using an actual biological system.

**ACKNOWLEDGMENTS**

This work was supported by a Grant-in-Aid from the Ministry of Education, Culture, Sports, Science, and Technology of Japan (*Synergy of Fluctuation and Structure: Quest for Universal Laws in Non-Equilibrium Systems,* P2013-201 Grant-in-Aid for Scientific Research on Innovative Areas, MEXT, Japan). There are no conflicts of interest.

**Figure 1**

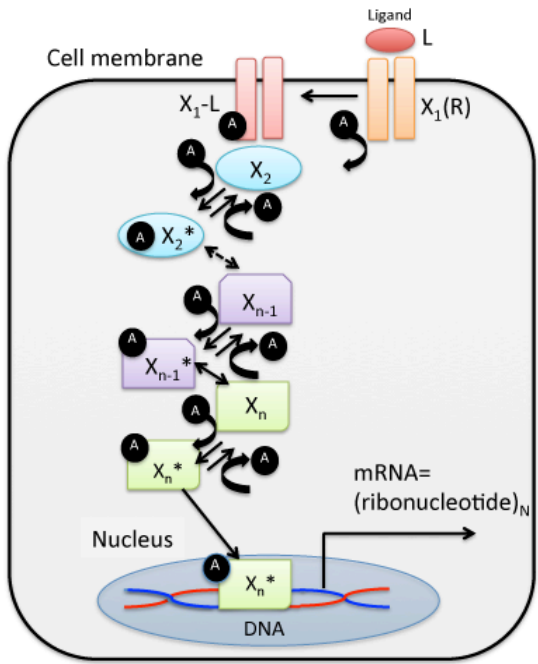

Figure 1. Schematic of a reaction cascade in cell signal transduction. *L* is a ligand and *R* is a receptor that mediates the cellular responses to external environmental changes. *A* is a messenger of signal transduction. Individual signaling molecules $X_j$ ($1 \leq j \leq n$) relay the modification of individual steps, and the last species $X_n$ is translocated to the nucleus, where it controls gene expression by transcription of mRNA.

**Figure 2**

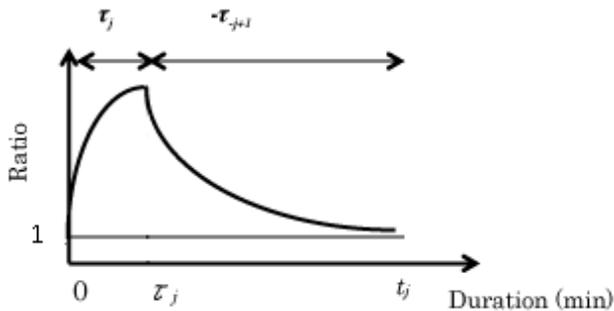

Figure 2. A common time course of the *j*-th to *j+1*-th step. Fold changes in modification. The vertical axis denotes the ratio $(X_j^* + \Delta X_j^*)/X_j^*$. The horizontal axis denotes the duration (min) of the $j^{th}$ step. $t_j$ represents the total duration of the $j^{th}$ step and is equivalent to $\tau_j - \tau_{-j+1}$. The suffix $-j+1$ implies that the $j+1^{th}$ molecule transmits opposite signaling to the $j^{th}$ molecule, for which the duration $\tau_{-j+1}$ has negative values.

Figure 3

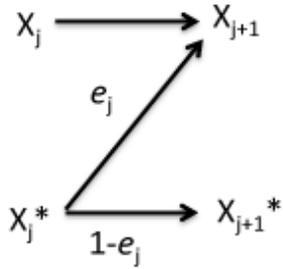

Figure 3. Schematic representation of the relationship between inputs and outputs at the *j*-th signaling step of a simple discrete channel. $e_j$ denotes the error probability in signal transduction from $X_j^*$ to $X_{j+1}$ instead of to $X_{j+1}^*$.

Figure 4

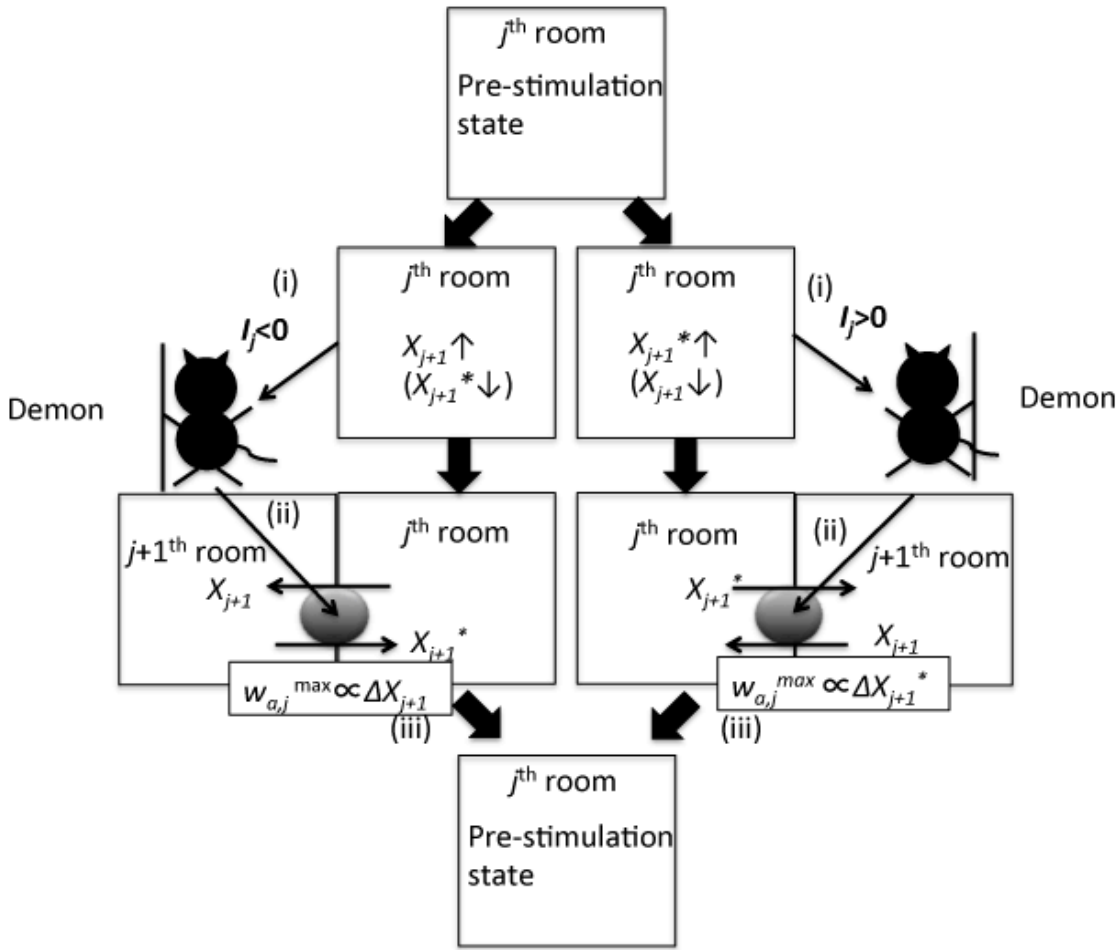

Figure 4. Feedback controller schematic. (i) The feedback controller observes whether $X_{j+1}^*$ increases or $X_{j+1}$ increases. (ii) The controller opens the gate for the increased $\Delta X_{j+1}^*$ or for the increased $\Delta X_{j+1}$ to enter the $j + 1^{th}$ room to prevent further proceeding of the signal. (iii)

The $j^{th}$ room in the system recovers to the initial state in this reaction cycle. The maximum chemical taken-out work $w_{a,j}^{max}$ from $j^{th}$ room may be described in proportion to $\Delta X_{j+1}*$ and $\Delta X_{j+1}$ (10-12). The feedback controller has the potential to recognize the concentration difference of the messenger between the initial state and final state at the $j^{th}$ room with information $\log A_{fj}/A_{ij}$ given in Eq. [57]. Subsequently, the controller feeds back the information into the work for the transfer of $X_{j+1}*$ from the $j^{th}$ room.

Figure 5

A

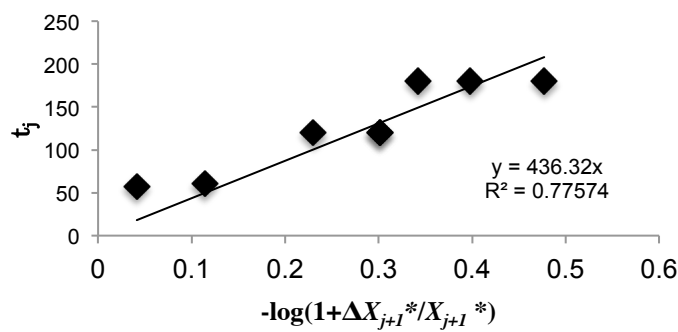

B

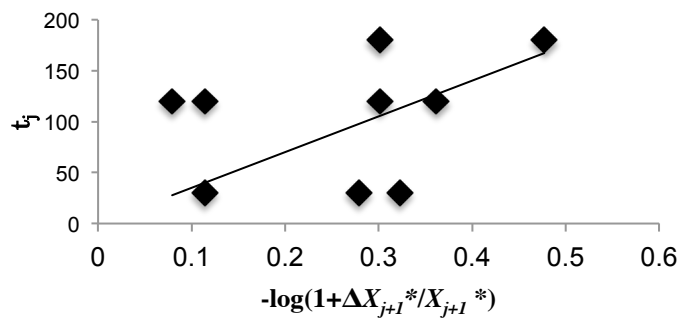

**Figure 5. Quantification of the MAPK signaling pathway.**
(A, B) A regression analysis of $-\log(1+\Delta X_{j+1}*/X_{j+1}^{st}*)$ was performed for signal duration ($t_j$) (min) in the RRE with three steps(19). Regression lines are illustrated in the plots. The gradient of the regression line represents the reciprocal of channel capacity $\Delta \zeta$.